\pdfoutput=1

\documentclass[11pt]{article}

\usepackage[review]{ACL2023}
\usepackage{amsmath}
\usepackage{graphicx}
\usepackage{times}
\usepackage{latexsym}
\usepackage{color}
\usepackage[T1]{fontenc}

\usepackage[utf8]{inputenc}
\usepackage{microtype}

\usepackage{inconsolata}

\title{MOSPC: MOS Prediction Based on Pairwise Comparison}

\author{Kexin Wang, Yunlong Zhao, Qianqian Dong, Tom Ko, Mingxuan Wang\\
        ByteDance, China \\ \texttt{\{wkx, zhaoyunlong.123, dongqianqian, tom.ko, wangmingxuan.89\}@bytedance.com}
}

\begin{document}
\maketitle

\begin{abstract}
    As a subjective metric to evaluate the quality of synthesized speech, Mean opinion score~(MOS) usually requires multiple annotators to score the same speech. Such an annotation approach requires a lot of manpower and is also time-consuming. MOS prediction model for automatic evaluation can significantly reduce labor cost. In previous works, it is difficult to accurately rank the quality of speech when the MOS scores are close. However, in practical applications, it is more important to correctly rank the quality of synthesis systems or sentences than simply predicting MOS scores. Meanwhile, as each annotator scores multiple audios during annotation, the score is probably a relative value based on the first or the first few speech scores given by the annotator. 
Motivated by the above two points, we propose a general framework for \textbf{MOS} prediction based on \textbf{p}air \textbf{c}omparison (MOSPC), and we utilize \textit{C-Mixup} algorithm to enhance the generalization performance of MOSPC.
The experiments on BVCC and VCC2018 show that our framework outperforms the baselines on most of the correlation coefficient metrics, especially on the metric KTAU related to quality ranking. And our framework also surpasses the strong baseline in ranking accuracy on each fine-grained segment. These results indicate that our framework contributes to improving the ranking accuracy of speech quality.
\end{abstract}

\section{Introduction}
\label{sec:intro}
Speech quality evaluation metrics are designed to reflect the speech quality of synthesized speech. Speech quality evaluation metrics include objective metrics~\cite{ref1,ref2,ref3} and subjective metrics~\cite{ref4}. 

MOS prediction is the task of constructing an automatic evaluation metric by fitting the subjective evaluation metric MOS. The training process of previous works mainly focus on predicting the MOS of a single speech. 
By reviewing the annotation process of MOS, we found that comparison may be a potential scoring strategy employed by some of the annotators. Specifically, in the dataset VCC2018, each annotator scored an average of 226 speech. As each annotator annotates multiple speech in succession, the scores given by some annotators may be relative scores after comparison (e.g., the first or first few utterances scored by the annotator may be used as a benchmark).
Moreover, compared with predicting the specific MOS score values of the speech samples, ranking the quality of speech samples has more practical application value and is often more difficult when speech samples have close MOS scores. Many previous works~\cite{ref7} have raised the problem of generalization, and the performance will be significantly degraded when facing the out-of-distribution~(OOD) problems. Motivated by the above points, we propose a \textbf{MOS} prediction model based on \textbf{p}airwise \textbf{c}omparison (\textbf{MOSPC}). Our contributions can be summarized as follows:

\begin{itemize}
	\item We propose a general framework for MOS prediction based on pair comparison, which forces the model to pay more attention to correctly rank the quality of two speech samples. To verify that MOSPC contributes to speech quality ranking, we test the ranking accuracy on the validation set on fine-grained MOS score segments. Then we utilize the C-Mixup algorithm to enhance the performance of generalization on BVCC. 
	\item Our proposed framework outperforms the baselines on BVCC and VCC2018 on most of the correlation coefficient metrics, especially on the metric KTAU related to quality ranking. And our framework surpasses the strong baseline in ranking accuracy on each fine-grained segment. These results indicate that our framework contributes to improving ranking accuracy. The model trained with VCC2018 and BVCC outperforms  baselines on the OOD datasets VCC2016 and BC2019 in zero-shot experiments respectively. 
    In addition, we analyze  the performance of our model for fine-grained OOD
 categories, such as unseen-system, unseen-listener and unseen-speaker.
\end{itemize}

\section{Related Work}
\label{sec:related}
A classic work in MOS prediction task is MOSNET~\cite{ref5}, which adopts the model structure of CNN-BiLSTM and proposes a loss function combining frame-level loss and utterance-level loss. Due to the need for manual annotation, few data can be used in the MOS prediction task. To reduce data waste, MBNET~\cite{ref6} proposes a MOS predictor consisting of a meanNet and a biasNet. 
LDNET~\cite{ref11} 
observed that MBNet removes biasNet at inference and only retains meanNet, which is inefficient. Therefore, LDNET improves MBNET by adopting an encoder-decoder structure to reduce the waste of parameters.
DDOS~\cite{ref12} proposes to eliminate the domain mismatch between self-supervised learning (ssl) model and MOS prediction data, and adds score distribution of each utterance to model learning. 
UTMOS~\cite{ref13} is based on ensemble learning of strong and weak learners. Fusion-SSL~\cite{ref14} uses late fusion, and fuses the results of $7$ ssl models to predict MOS value.
\citet{ref7} makes a analysis of the OOD problem of MOS prediction. The OOD problems in MOS prediction mainly include unseen-system, unseen-listener, unseen-speaker in the test and validation sets.

Our proposed MOSPC adopts dynamic pairwise comparison. Compared with the previous methods~\cite{ref5,ref6,ref11,ref14}, our method pays more attention to correctly evaluating the relative quality of speech. 

\section{Method}
\label{sec:method}
In this section, we will introduce the overall structure of MOSPC and the implementation of pairwise comparison, as well as the C-Mixup algorithm used to enhance generalization performance.
\subsection{Preliminary}
Given a dataset $D$ including $N$ speech samples, denoted as $D=\{[x_1,y_1],[x_2,y_2], \ldots ,[x_N,y_N]\}$, $x_i$ and $y_i$ denote the $i$th speech sample and its ground truth.
We denote the $k$th ssl model as $f_k$, $k\in \{1,2,\ldots,7\}$, then the predicted MOS of the $k$th ssl model for input $x_i$ can be represented as $m_{ki}=f_k(x_i)$. $F$ represents the fusion model, which consists of $7$ ssl models and a fusion layer. $m_i=F(x_i)$ denotes the predicted MOS made by the fusion model.
\begin{table*}[t]
    \caption{Results on VCC2018 and BVCC. The left side of the table shows the results of our proposed MOSPC and baselines on VCC2018. The right side of the table shows the results of our proposed MOSPC and baselines on BVCC.}
    \centering
    \small
    \setlength{\tabcolsep}{0.6mm}{
    \begin{tabular}{c| c c c c | c c c c | c c c c | c c c c}
        \hline
        &\multicolumn{8}{c}{VCC2018} & \multicolumn{8}{|c}{BVCC}\\
        \hline
         & \multicolumn{4}{c|}{utterance-level} & \multicolumn{4}{c}{system-level} & \multicolumn{4}{|c|}{utterance-level} & \multicolumn{4}{c}{system-level}\\
         \hline
          & MSE & LCC & SRCC & KTAU & MSE & LCC & SRCC & KTAU & MSE & LCC & SRCC & KTAU & MSE & LCC & SRCC & KTAU \\
          \hline
          MOSNET & 0.538 & 0.643 & 0.589 & - & 0.084 & 0.957 & 0.888 & - & 0.816 & 0.294 & 0.263 & - & 0.563 & 0.261 & 0.266 & -  \\
          \hline
          LDNET & 0.441 & 0.664 & 0.626 & 0.465 & 0.022 & 0.978 & 0.932 & 0.825  & 0.338 & 0.774 & 0.773 & 0.582 & 0.139 & 0.896 & 0.893 & 0.714\\
          \hline
          MBNET & 0.426 & 0.680 & 0.647 & - & 0.029 & 0.977 & 0.949 & - & 0.433 & 0.727 & 0.753 & 0.564 & 0.228 & 0.844 & 0.870 & 0.685 \\
          \hline
          Fusion-SSL & 0.359 & 0.740 & 0.711 & 0.542 & \textbf{0.018} & 0.991 & 0.984 & 0.914 & 0.156 & 0.902 & 0.901 & 0.735 & \textbf{0.051} & \textbf{0.960} & \textbf{0.962} & \textbf{0.848}\\
          \hline
          MOSPC & \textbf{0.352} & \textbf{0.748} & \textbf{0.721} & \textbf{0.551} & 0.020 & \textbf{0.993} & \textbf{0.988} & \textbf{0.938}  & \textbf{0.148} & \textbf{0.906} & \textbf{0.906} & \textbf{0.742} & 0.054 & \textbf{0.960} & \textbf{0.962} & 0.841\\
          \hline
    \end{tabular}
    }
    \label{main track}
\end{table*}
\subsection{MOSPC}
\subsubsection{Fusion Model}
\begin{figure}
\centerline{\includegraphics[width=\columnwidth]{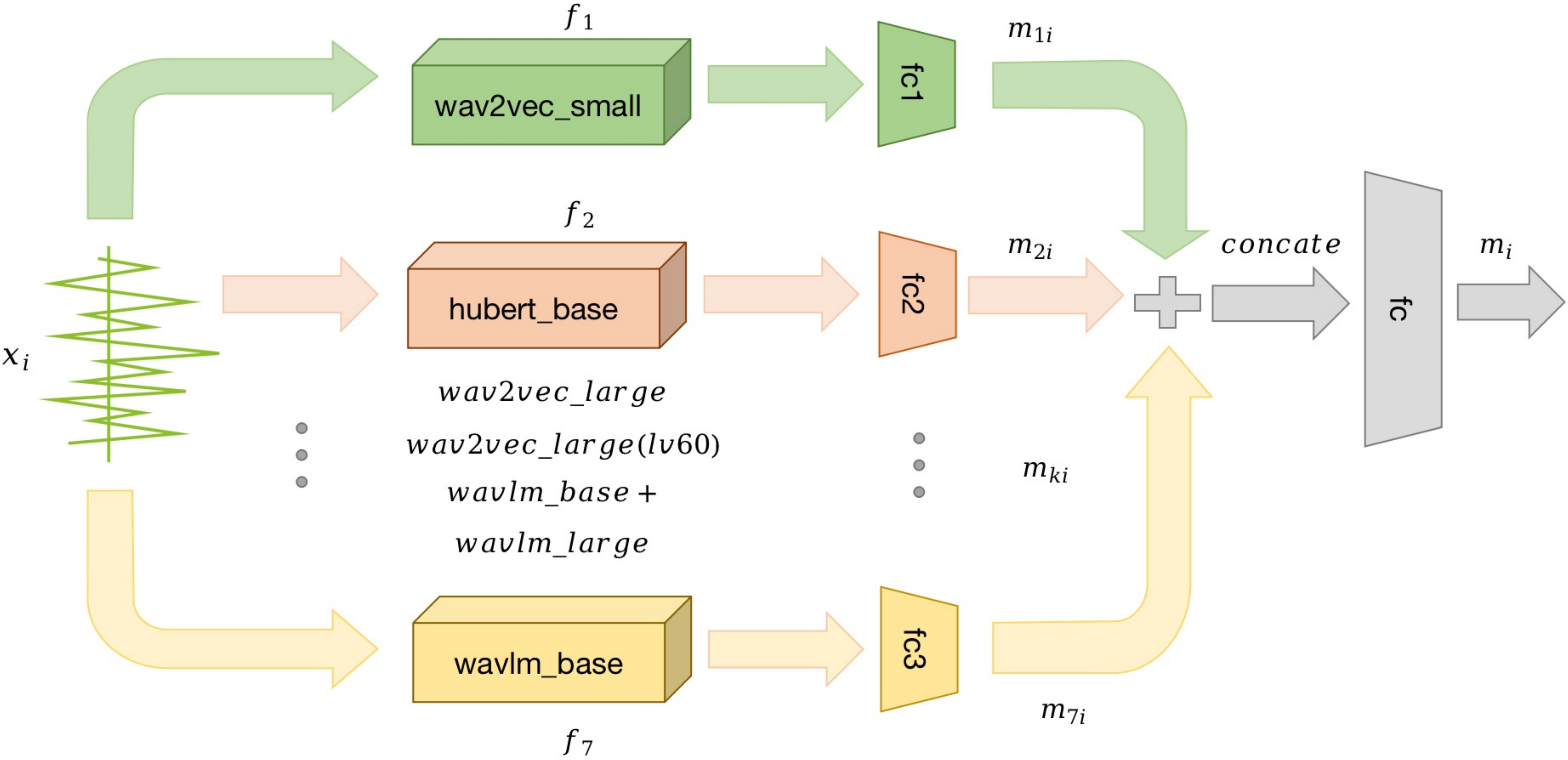}}
\caption{Overall model structure for inference. 
}
\label{whole model}
\end{figure}

Our model is based on Fusion-SSL~\cite{ref14}. The overall model structure is shown in Figure \ref{whole model}. The fusion model mainly consists of 7 ssl models: \textit{wav2vec\_small}, \textit{wav2vec\_large}, \textit{hubert\_base}, \textit{wav2vec\_large(lv60)}, \textit{wavlm\_base}, \textit{wavlm\_base+}, \textit{wavlm\_large} and a fusion layer. The fusion layer is a fully connected layer. During inference, speech $x_i$ is fed to ssl model $f_1,f_2,\ldots,f_7$ separately, and the MOS values $m_{1i},m_{2i},\ldots,m_{7i}$ are obtained. Then the MOS values are concatenated and fed into the fusion layer to predict MOS value $m_i$. During training, we leverage pairwise comparison to force the model to pay more attention to the relative quality of speech.

\begin{figure}
\centerline{\includegraphics[width=\columnwidth]{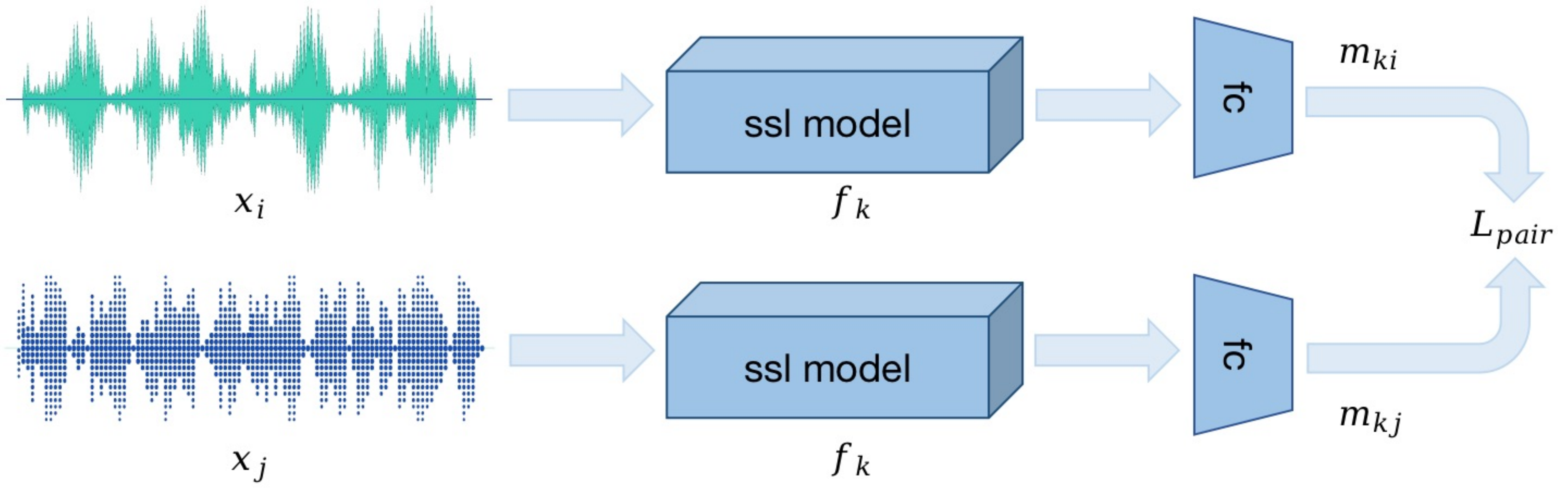}}
\caption{Training process based on pair comparison. 
}
\label{train}
\end{figure}
\subsubsection{Training in Stages}
\paragraph{Pair Comparison} Our proposed training process is shown in Figure \ref{train}. We dynamically make pairs in each batch and constrain each speech sample to form at most two pairs in order to prevent overfitting. The speech samples $x_i$ and $x_j$ are input into the ssl model respectively, then MOS scores $m_{ki}$ and $m_{kj}$ are predicted, and the loss function $L_{pair}$ is calculated jointly. All $7$ ssl models are trained in such a pair comparison manner.  Loss function $L_{pair}$ consists of three parts: the relative ranking loss $L_{rank}$ and the L1 loss of two speech samples denoted by $L_{d1}$ and $L_{d2}$ respectively:
\begin{align}
    L_{pair}=(1-\beta)*L_{rank}+\beta*(L_{d1}+L_{d2})
\end{align}
where $\beta$ is a hyperparameter, the model learns to predict MOS scores by optimizing $L_{d1}$,$L_{d2}$, and learns to rank two audios by optimizing $L_{rank}$~\cite{ref15}. Refer to Appendix A for more details of $L_{rank}$.

\paragraph{C-Mixup} We observed degradation in generalization performance in the experiments on the BVCC dataset. Therefore, after experiments in Section \ref{subsec:mos} and \ref{subsec:rank}, for each ssl model trained on BVCC, we adopt C-Mixup~\cite{ref16,cheng2023m} to enhance the generalization performance. C-Mixup proportionally combines in-set samples in pairs to construct pseudo-out-of-set samples to improve the generalization performance of the model. Refer to Appendix B for details of C-Mixup algorithm. To distinguish the model from the one trained without C-Mixup, we named the model trained with C-Mixup as \textbf{MOSPC-C}.

\section{Dataset}
\label{sec:dataset}

The datasets adopted in this paper include main track data and out-of-domain data. Main track data include VCC2018 and BVCC, and out-of-domain data include VCC2016, BC2019 and ASV2019. See Appendix C for details of datasets. 

\section{Experiments and Discussion}
\label{sec:experimet}
In this section, we will compare the performance of MOSPC with the baselines~\cite{ref5, ref6, ref11} and strong baseline Fusion-SSL~\cite{ref14} on the datasets BVCC and VCC2018. We test the generalization performance on BC2019 and VCC2016. We also list the ranking accuracy of fine-grained MOS segments.

\subsection{Experiment Settings}
We leverage a fusion layer and $7$ ssl models
to form the overall model. Each ssl model was trained in the pair comparison manner with SGD optimizer for 1000 epochs. We applied early stopping based on the L1 loss of the validation set with 20 epochs patience, and set the learning rate to 1e-4, batch size of 8. The hyperparameter $\beta$ was set to be 0.6. After training on the BVCC in a pair comparison manner, we also use the C-Mixup algorithm to enhance the generalization performance of the model. 
When trained with the C-Mixup algorithm, We set the bandwidth to be 1.0, $\alpha$ to be 2.0. We implemented our models in Fairseq~\cite{ott2019fairseq}. All experiments were performed on 7 32GB GPUs.

\subsection{MOS Prediction Results}
\label{subsec:mos}
The left side of Table \ref{main track} shows the results of MOSPC and baselines on VCC2018. MOSPC outperforms baselines in all correlation coefficient metrics of utterance-level. At the system-level, all correlation coefficient metrics outperform baselines except the MSE, which was slightly higher by 0.002. The remarkable thing is that the KTAU of system-level surpasses the baselines significantly. KTAU is a correlation coefficient metric used to indicate the ranking correlation between the predicted value and the ground truth. These results indicate that our framework contributes to the ranking correctness improvement, which is in line with our motivation.


The right side of Table \ref{main track} shows the results of our proposed MOSPC and baselines on BVCC. The results show that our model outperforms previous works on all correlation coefficient metrics at utterance-level, especially on KTAU. At the system-level our framework matches the strong baseline performance on LCC and SRCC. As there are unseen-system samples in the BVCC validation set, the performance of system-level will be affected by the samples from unseen systems. These results also imply that the pair-wise training may impair the generalization performance. To solve this problem, we adopted C-Mixup algorithm to improve the generalization performance of our model trained on BVCC.

\subsection{Ranking Accuracy on Fine-Grained MOS Segments}
\label{subsec:rank}
To prove that our proposed framework contributes to the improvement of speech ranking accuracy, we analyzed the ranking accuracy of speech quality on fine-grained MOS score segments. As shown in Table \ref{fine-grain}, on BVCC and VCC2018, we divided the ground truth 1-5 into four segments with a score interval of 1. 
On each fine-grained MOS segments and the overall MOS segment 1-5, we calculated the ranking accuracy on each speech pair $(x_i,x_j)$ with ground truth $\lvert y_i-y_j \rvert \in (0,1]$. This is because, from our motivation, we pay more attention to whether our proposed framework can accurately rank speech with different but close score values. 

\begin{table}[th]
    \centering
    \caption{Ranking accuracy on fine-grained MOS segments. "1-2","2-3","3-4" and "4-5" are the fine-grained segments and "1-5" is the overall segment.
    }
    \setlength{\tabcolsep}{1mm}{
    \begin{tabular}{c | c | c | c | c | c}
    \hline
    \multicolumn{6}{c}{BVCC}\\
    \hline
    & 1-2 & 2-3 & 3-4 & 4-5 & 1-5\\
    \hline
    Fusion-SSL & 0.728 & 0.724 & 0.739 & 0.675 & 0.778\\
    \hline
    MOSPC & \textbf{0.731} & \textbf{0.737} & \textbf{0.742} & \textbf{0.679} & \textbf{0.787} \\
    \hline
    \multicolumn{6}{c}{VCC2018}\\
    \hline
    Fusion-SSL & 0.482 & 0.469 & 0.515 & 0.509 & 0.493\\
    \hline
    MOSPC & \textbf{0.489} & \textbf{0.473} & \textbf{0.517} & \textbf{0.514} & \textbf{0.494} \\
    \hline
    \end{tabular}
    }
    \label{fine-grain}
\end{table}

The top half of Table \ref{fine-grain} shows the fine-grained MOS segments ranking results on the validation set of BVCC. The bottom half of Table \ref{fine-grain} shows the fine-grained 
MOS segments ranking results on the validation set of VCC2018. The result shows that our proposed framework outperforms the strong baseline in ranking accuracy on each segment on both BVCC and VCC2018. These results indicate that our framework contributes to improving the ranking accuracy of speech samples with different but close MOS scores.

\subsection{OOD Experiments}
We first analyze the generalization performance of models trained with VCC2018 on VCC2016. As shown in Table \ref{VCC2016}, since VCC2016 only has system-level labels, we only present the system-level results. Our proposed framework outperforms previous works in all metrics, and the improvement is also significant in the KTAU metric, which again proves that our proposed framework contributes to correctly ranking the relative quality of speech.

\begin{table}[th]
    \centering
    \caption{Zero-shot experiment results on VCC2016.}
    \setlength{\tabcolsep}{1.4mm}{
    \begin{tabular}{c|c|c|c|c}
    \hline
    \multicolumn{5}{c}{VCC2016}\\
    \hline
    & \multicolumn{4}{c}{system-level}\\
    \hline
    & MSE& LCC & SRCC & KTAU\\
    \hline
      MBNET& 0.207 & 0.931 & 0.906  &  -\\
      LDNET &0.215 & 0.939 & 0.896  & 0.768\\
      Fusion-SSL &0.209 & 0.971 & 0.889 & 0.768 \\
      MOSPC &\textbf{0.121} & \textbf{0.983} & \textbf{0.935} & \textbf{0.832}\\
      \hline
    \end{tabular}
    }
    \label{VCC2016}
\end{table}
As mentioned before, from the experiments on the BVCC validation set we found that the pair-wise training method may lead to a decrease in generalization performance, so we leveraged the C-Mixup algorithm to improve the generalization performance on the BVCC after experiments in section \ref{subsec:mos} and \ref{subsec:rank}. Table \ref{bc2019} lists the zero-shot results of  BC2019. The zero-shot results indicate that after training with C-Mixup, the generalization performance improved significantly, and the robustness to the unseen-system and multi-languages OOD challenges is also improved.

\begin{table}[h]
    \centering
    \caption{Zero-shot experiment results on BC2019. MOSPC-C indicates the model trained with C-Mixup algorithm}
    \label{bc2019}
    \setlength{\tabcolsep}{0.1mm}{
    \begin{tabular}{c|c c c|c c c}
         \hline
        \multicolumn{7}{c}{BC2019}\\
         \hline
         & \multicolumn{3}{c|}{utterance-level} & \multicolumn{3}{c}{system-level}\\
         \hline
         & LCC & SRCC & KTAU & LCC & SRCC & KTAU\\
         \hline
         LDNET & 0.384 & 0.365 & 0.252 & 0.500 & 0.473 & 0.354  \\
         \hline
         DDOS & 0.678 & 0.694 & 0.502 & 0.766 & 0.797 & 0.637 \\
         \hline
         Fusion-SSL & 0.718 & 0.642 & 0.469 & 0.803 & 0.792 & 0.601 \\
         \hline
         MOSPC & 0.704 & 0.709 & \textbf{0.523} 
         & 0.731 & 0.778 & 0.594 \\
         \hline
         MOSPC-C & \textbf{0.756} & \textbf{0.711} & 0.521 & \textbf{0.816} & \textbf{0.851} & \textbf{0.667} \\
        \hline
    \end{tabular}
    }
\end{table}
\begin{table*}[h]
    \centering
    \caption{Analysis of fine-grained OOD catagories. Mean and standard deviations of squared errors for fine-grained OOD catagories of unseen-speaker, unseen-system and unseen-listener are shown.}
    \label{unseen}
    \setlength{\tabcolsep}{5mm}{
    \begin{tabular}{c|c|c|c|c}
         \hline
        & unseen-speaker &\multicolumn{2}{c|}{unseen-system}&unseen-listener\\
         \hline
         & ASV2019 & ASV2019 & BC2019 & ASV2019\\
         \hline
         Fusion-ssl & 1.104±1.641 & 1.114±1.707 & 0.191±0.225 & 1.032±1.558\\
         \hline
         MOSPC & 1.098±1.602 & 1.124±1.690 & 0.189±0.213 & 1.041±1.572\\
         \hline
         MOSPC-C & \textbf{1.089±1.587} & \textbf{1.103±1.673} & \textbf{0.179±0.217} & \textbf{1.030±1.547}\\
         \hline
    \end{tabular}
    }
\end{table*}

We followed \cite{ref7} analyzing the performance of our model on the fine-grained OOD categories of unseen-system, unseen-listener and unseen-speaker with ASV2019 and BC2019. We first adopted ASV2019 and BC2019 to fine-tune the model trained on BVCC respectively. As shown in table \ref{unseen}, we report the mean and standard deviations of squared errors for the unseen categories on utterance-level. The results indicate that our proposed method performs better on the category unseen-listener than on unseen-speaker and unseen-system.

\section{Conclusion}
This paper proposes a general framework for MOS prediction based on pairwise comparisons~(MOSPC) to solve the problem that it is difficult for MOS prediction models to correctly rank speech quality when the MOS scores are close. The main track experiment results show that MOSPC outperforms baselines on most of the correlation coefficient metrics, especially on the metric KTAU related to speech quality ranking.
Moreover, MOSPC surpasses the strong baseline in ranking accuracy on each fine-grained segment. These results indicate that training in a pair comparison manner contributes to improving ranking accuracy. We leverage C-Mixup algorithm to enhance the generalization performance. On the OOD datasets VCC2016 and BC2019, our method outperforms baselines on all metrics. We also analyze the performance on fine-grained OOD categories. Our method performs better for the unseen-listener OOD category than for the unseen-speaker and unseen-system OOD categories.

\section{Limitation}
MOSPC can improve ranking accuracy on each fine-grained MOS score segment, but at the same time, the training method based on pair comparison may impair the generalization performance. As there are unseen-systems in the BVCC validation set, the system-level results of BVCC are affected by the generalization performance degradation. We introduced the C-Mixup algorithm to enhance the generalization performance, which increased the complexity of the experiment to some extent.

\bibliography{anthology,acl2023}
\bibliographystyle{acl_natbib}

\appendix

\label{sec:appendix}
\section{Details of the Relative Ranking Loss}
\label{sec:appendix}
$L_{rank}$ was introduced by rankNet\cite{ref15}. $L_{rank}$ is similar in form to cross entropy loss:
\begin{align}
    L_{rank}=-L*log(P)-(1-L)*log(1-P)
\end{align}
where $L_{rank}$ maps the outputs $m_{ki}$, $m_{kj}$ into probability $P$ via a logistic function:
\begin{align}
    P=\frac{e^{m_{ki}-m_{kj}}}{1+e^{m_{ki}-m_{kj}}}
\end{align}
The value of $L$ depends on the ground truths of two speech samples.
\begin{align}
    L=\left\{\begin{aligned}0, & & y_i<y_j\\0.5, & & y_i=y_j\\1, & & y_i>y_j\end{aligned}\right.
\end{align}

\section{C-Mixup}
\begin{figure}
\centerline{\includegraphics[width=\columnwidth]{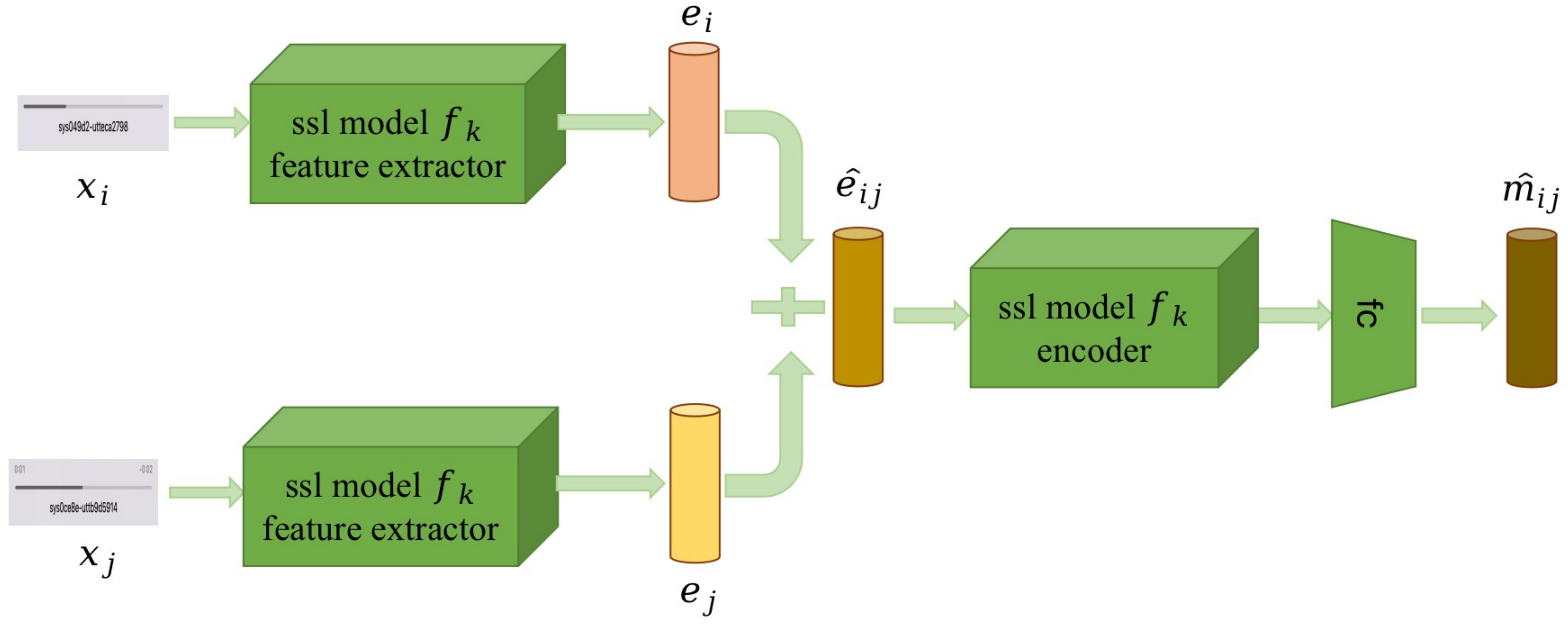}}
\caption{Illustration of the training process of C-Mixup. 
}
\label{cmixup}
\vspace{-1.5em}
\end{figure}

For ssl model $f_k$ and input speech sample $(x_i,y_i)$, we need to sample another instance $(x_j,y_j)$ from the training set. C-Mixup first constructs a sampling probability distribution based on a symmetric Gaussian kernel for each audio sample $x_i$:
\begin{align}
    P((x_j,y_j)\mid (x_i,y_i))\propto exp(-\frac{d(i,j)}{2\sigma^2})
\end{align}
where $d(i,j)=d(y_i,y_j)=\Vert y_i-y_j \Vert_2^2$ represents the distance between $y_i$ and $y_j$, and $\sigma$ represents the bandwidth which is a hyperparameter. Subsequently, these conditional probabilities are normalized into a probability mass function that sums to one, and another sample is selected by sampling through the probability mass function. Figure \ref{cmixup} illustrates the training process of C-Mixup. Each ssl model in this work contains two parts: feature extractor and encoder. $x_i$ and $x_j$ are fed into the feature extractor respectively to obtain embedding $e_i$ and $e_j$. Then $e_i$ and $e_j$ are proportionally  combined to construct the embedding $\hat{e}_{ij}$ of pseudo-out-of-set sample:
\begin{align}
    \hat{e}_{ij}=\lambda*e_i+(1-\lambda)*e_j
\end{align}
where $\lambda\sim Beta(\alpha,\alpha)$, and $\alpha$ is the parameter of the Beta distribution. $\alpha$ is a hyperparameter. The remaining models take the pseudo-out-of-set embedding $\hat{e}_{ij}$ as input to predict MOS score $\hat{m}_{ij}$, and compute the L1 loss with $\hat{y}_{ij}$. $\hat{y}_{ij}$ is constructed in the same way as $\hat{e}_{ij}$:
\begin{align}
    \hat{y}_{ij}=\lambda*y_i+(1-\lambda)*y_j
\end{align}
consistent with the main track, each ssl model is trained with C-Mixup algorithm separately.

\section{Details of Dataset}
\subsection{Main Track Data}
\subsubsection{VCC2018}
Samples in VCC2018 were all sampled from Voice Conversion Challenge 2018\cite{ref17}, including $20580$ English speech samples synthesized by $38$ systems. A total of $267$ professional annotators participated in the speech labeling. Each speech was scored by four annotators, and the four integer scores were averaged as the label. For the sake of comparison, We split VCC2018 into training sets with a size of $13580$, validation sets with a size of $3000$ and test sets with a size of $4000$.
\subsubsection{BVCC} 
BVCC integrates data from multiple synthetic speech competitions, including Blizzard Challenge\cite{ref18, ref19, ref20, ref21, ref22, ref23}, the Voice Conversion Challenge\cite{ref24,ref25,ref26,ref27,ref28} and publicly-available samples from systems implemented in ESPnet\cite{ref29}. 
BVCC includes a total of 7106 English speech samples submitted by 187 systems. We split BVCC into training, validation and test sets with a rate of $70\%$, $15\%$ and $15\%$. Each speech was scored by eight annotators, and the eight integer scores were averaged as the label. Unlike VCC2018, BVCC has samples from unseen systems in its validation set.
\subsection{Out-of-domain Data}
\subsubsection{VCC2016}
In order to compare with previous works, we adopt VCC2016 to test the OOD performance of models trained with VCC2018. VCC2016 includes $26028$ speech samples synthesized by $20$ systems. VCC2016 has only system-level labels and without utterance-level labels. 
\subsubsection{BC2019}
We adopt BC2019 to test the OOD performance of models trained with BVCC. Samples in BC2019 are all sampled from Blizzard Challenge 2019, and are Chinese TTS synthesized speech rated by Chinese native speakers. 
Since all samples of BVCC are in English, BC2019 can be used as a cross-language OOD case to test the generalization performance of models trained with BVCC. BC2019 has provided 136 labeled samples for training, 136 samples for validation, and additional 540 unlabeled data.

\subsection{ASV2019}
We follow \cite{ref7} utilizing ASV2019\cite{WANG2020101114,Todisco2019} to analyze the performance of our model on fine-grained OOD experiments. Samples in ASV2019 are all in English and sampled from the human assessment results data on the ASVspoof2019 database LA scenario. As scores in human assessment results data are distributed from 0 to 9, We linearly project the scores to 1-5.

\end{document}